\documentclass[a4paper,11pt]{article}
\usepackage{diagbox}
\usepackage{graphicx}
\usepackage{here}
\usepackage{amsmath,amssymb,amsthm}
\usepackage{mathrsfs}
\usepackage{ascmac}
\usepackage[sort, comma]{natbib}
\usepackage[subrefformat=parens]{subcaption}
\usepackage{url}
\usepackage{tikz}
\usetikzlibrary{intersections,calc}

\newtheorem{theo}{Theorem}
\newcommand{\ol}[1]{\overline{#1}}

\title{
\begin{flushleft}
{\Large
Agglomeration triggered by the effect of the number of regions: A model in NEG with a quadratic subutility}
\end{flushleft}
}

\author{Kensuke Ohtake\thanks{Center for Mathematical Modeling and Data Science, Osaka University, Toyonaka, Osaka 560-8531, Japan,
E-mail: k-ohtake@sigmath.es.osaka-u.ac.jp}}
\date{March 23, 2022}

\begin{document}
\maketitle

\begin{abstract}
We extend the mathematical model proposed by Ottaviano-Tabuchi-Thisse (2002) to a multi-regional case and investigate the stability of the homogeneous stationary solution of the model in a one-dimensional periodic space. When the number of regions is two and three, the homogeneous stationary solution is stable under sufficiently high transport cost. On the other hand, when the number of regions is a multiple of four, the homogeneous stationary solution is unstable under any values of the transport cost.
\end{abstract} 

\noindent
{\bf Keywords\hspace{3mm}}new economic geography~\textperiodcentered~differential equations~\textperiodcentered~racetrack economy~\textperiodcentered~self-organization~\textperiodcentered~spatial patterns~\textperiodcentered~number of cities


\section{Introduction}
In considering spatial economy, how the spatial structure formed by economic factors (such as labor and capital) emerges has been a major theoretical concern. Mathematically, the occurrence of such emergence of spatial structure corresponds to the instability of the homogeneous (which means spatially uniform in the current context) stationary solution of a mathematical model in study. Therefore, it is significant to pay attention to unstable solutions of the model.  In our study, we work with a mathematical model that has been developed in new economic geography (NEG).

In NEG, the core-periphery model by \citet{Krug91} is one of the standard models in the field, and various applied studies have been conducted based on this model. The core-periphery model is very important in that it facilitates the treatment of spatial economic phenomena in a general equilibrium framework, but on the other hand, it lacks some realistic factors such as price competition among firms and the transport costs that depend on the volume of transportation (\citet[Chapter 8]{ZenTaka}). \citet{OttaTabThi} have proposed a mathematical model that incorporates the above factors and is easy to handle analytically by introducing a quasi-linear utility function of consumers. The approach of \citet{OttaTabThi} is considered to be a complementary approach to modeling that relies on the CES function such as the core-periphery model (\citet[Section 9.1]{FujiThis}). There are various theoretical studies based on models that use the quasi-linear utility functions similar to \citet{OttaTabThi} (See for example \citet{PicaZeng05}, \citet{Zeng06}, \citet{TakaZeng13}). The properties of the quasi-linear utility function have also been investigated in detail in the field of industrial organization (See, for example, \citet{SiVi} and \citet{AmErJi}).

The model by \citet{OttaTabThi} (let us call it {\it OTT model} in short) assumes a two-region economy, and it shows that the symmetric equilibrium is stable when the transport cost is sufficiently high, and becomes unstable when the transport cost falls below a certain value. As is well known, this property is common to the basic models in new economic geography, such as the core-periphery model.

One would expect this property to hold in a multi-regional version of the OTT model. In this paper, we show that this expectation is true up to three-regional case, but is no longer true for four-regional case.  More precisely, when the small perturbations around a homogeneous stationary solution are decomposed into the Fourier series, all effective eigenfunctions are stable for sufficiently high transport cost if the number of regions is up to three. On the other hand, when the number of regions reaches four, an eigenfunction that are unstable for any values of the transport cost appears for the first time. More generally, we show that such an always unstable eigenfunction appears when the number of regions is a multiple of four. Meanwhile, it is difficult to make a general statement when the number of regions is five or more and not a multiple of four. At least, we can argue that the appearance of the always unstable eigenfunction of five regional model depends on the population of immobile workers.

The rest of the paper is organized as follows. Section 2 derives the model. Section 3 discusses a homogeneous stationary solution and investigates the stability of it. Section 4 concludes. Section 5 is an appendix.

\section{Model equations}
This section derives the mathematical model we handle. We first review the use of the Dixit-Stiglitz framework \citep{DS77} by \citet{OttaTabThi}, and then apply it to modeling multi-regional spatial economy.

\subsection{Dixit-Stiglitz framework}
As in \citet{OttaTabThi}, we assume the following utility function for consumers.
\begin{equation}\label{utility}
U = \alpha\int_0^n q(i)di -\frac{\beta-\gamma}{2}\int_0^n q(i)^2di-\frac{\gamma}{2}\left[\int_0^n q(i)di\right]^2 + q_0.
\end{equation}
Here, $q(i)$ and $q_0$ are the quantity of variety $i\in[0, n]$ of the manufacturing goods, and the quantity of the agricultural good which is num\'{e}raire, respectively. All the parameters are positive; $\alpha>0$, $\beta>0$, and $\gamma>0$. Then, $\alpha>0$ stands for the intensity of preference for the differentiated product. It is assumed that $\beta>\gamma$, which means that the consumers prefer to consume a greater variety of goods. Each consumer faces the following budget constraint.
\begin{equation}\label{budget}
\int_0^n p(i)q(i)di + q_0 = Y + \ol{q}_0,
\end{equation}
where $Y$ is the income of the consumer, and $\ol{q}_0$ is the initial endowment of the agricultural good. Here, $\ol{q}_0$ is assumed to be sufficiently large for the individual consumption of the agricultural good to be positive in market equilibrium (\citet[p.414]{OttaTabThi}). Maximizing \eqref{utility} under \eqref{budget} yields the optimal consumption of the variety $i$ as
\begin{equation}\label{DS_demand}
q(i) = a - bp(i) + c\int_0^n [p(j)-p(i)]dj.
\end{equation}
The indirect utility of the consumer is given by
\begin{equation}\label{indu}
V=S+Y+\ol{q}_0,
\end{equation}
where $S$ stands for the consumer surplus given by
\begin{equation}\label{CS}
S=\frac{a^2n}{2b}-a\int_0^n p(i)di+\frac{b+cn}{2}\int_0^n p(i)^2di - \frac{c}{2}\left[\int_0^n p(i)di\right]^2.
\end{equation}
\subsection{Spatial modeling}
The economy consists of $R \in \mathbb{N}$ discretely countable regions. These regions are represented by indices such as $x$, $y$ or $z\in\left\{0,1,2,\cdots,R-1\right\}$. Let $\phi_x\geq 0$ and $\lambda_x\geq 0$ denote the immobile and mobile workers' population in region $x$, respectively. Let the total amount of each population be $\Phi\geq 0$ and $\Lambda\geq 0$, i.e., 
\[
\sum_{x=0}^{R-1}\phi_x = \Phi,~~\sum_{x=0}^{R-1}\lambda_x = \Lambda.
\]

Let us denote the price in region $y$ of the variety produced in region $x$ by $p_{xy}$, and the demand in region $y$ for the variety produced in region $x$ by $q_{xy}$. By \eqref{DS_demand}, for any regions $x$ and $y$, the demand $q_{xy}$ can be represented as 
\begin{equation}\label{qxyor}
q_{xy} = a- (b + cn)p_{xy} + cG_y,
\end{equation}
where $G_y$ stands for price index in region $y$ given by $G_y = \sum_{z=0}^{R-1} n_z p_{zy}$. Here, $n_y$ denotes the number of manufacturing firms in region $y$ and is assumed to be equal to the number of the varieties of the manufacturing goods in the region.

Let us consider the profit-maximizing behavior of the manufacturing firms. Each firm is assumed to be engaged in the production of one variety of manufacturing goods. Thus, there are $n$ firms in the whole economy with equal number of the varieties. The number of the firms in region $x$ is denoted by $n_x$. The firm is supposed to pay a transport cost for each unit of product sold. The transport cost of transporting a unit of product from region $x$ to $y$ is denoted by $\tau|x-y|$, where $\tau>0$ and $|x-y|$ is the distance between the two regions defined in some sense\footnote{We explicitly define the distance function $|x-y|$ in a more specific setting later.}. It is assumed that $F$ units of mobile workers are needed as the fixed input, so the number of the firms in region $x$ is expressed by
\begin{equation}\label{nxbylamxF}
n_x = \frac{\lambda_x}{F}.
\end{equation}
It follows immediately from \eqref{nxbylamxF} that 
\begin{equation}\label{nbyLamF}
n = \frac{\Lambda}{F}.
\end{equation}
The nominal wage of the mobile workers in region $x$ is denoted by $w_x$. Then, the profit earned by each firm in region $x$ is given by
\begin{equation}\label{profitinx}
\begin{aligned}
\Pi_x = \sum_{z=0}^{R-1} \left(p_{xz} - \tau|x-z|\right)q_{xz}\left(\phi_z+\lambda_z\right) - Fw_x.
\end{aligned}
\end{equation}
Each firm in region $x$ sets price of its product to maximize the profit \eqref{profitinx}, assuming that the price index $G_y$ in \eqref{qxyor} is given. The first-order condition of optimality is
\[
\frac{\partial }{\partial p_{xy}} \Pi_{x} = 0.
\]
for $y=0,1,\cdots,R-1$. It yields 
\begin{equation}\label{pxyqxydvbcntauxy}
p_{xy} = \frac{q_{xy}}{b+cn} + \tau|x-y|. 
\end{equation}
for any regions $x$ and $y$. Moreover substituting \eqref{qxyor} into \eqref{pxyqxydvbcntauxy} and using \eqref{nxbylamxF} and \eqref{nbyLamF}, we obtain
\begin{align}
p_{xy} = \frac{a}{2(b+cn)} + \frac{c}{2(bF+c\Lambda)} \sum_{z=0}^{R-1} \lambda_z p_{zy} + \frac{\tau}{2}|x-y|. \label{prices} 
\end{align}
for any regions $x$ and $y$.

Let us derive an equation for the nominal wage. Under the free entry, the profit $\Pi_x=0$ in any region $x$. That is, 
\[
\sum_{z=0}^{R-1} \left(p_{xz} - \tau|x-z|\right)q_{xz}\left(\phi_z+\lambda_z\right) - Fw_x=0
\]
Therefore, 
\begin{equation}\label{wx}
w_x = \frac{1}{F}\sum_{z=0}^{R-1} \left(p_{xz} - \tau|x-z|\right)q_{xz}\left(\phi_z+\lambda_z\right).
\end{equation}
holds in each region $x$. It follows from \eqref{prices} that
\begin{equation}\label{pxyminustaudv2equal}
\left[p_{xy}-\frac{\tau}{2}|x-y|\right]2(b+cn) = a + \frac{c}{F} \sum_{z=0}^{R-1} \lambda_z p_{zy}.
\end{equation}
Combining \eqref{pxyminustaudv2equal} with \eqref{qxyor} yields
\begin{equation}\label{qxybcnpxymtauxy}
\begin{aligned}
q_{xy} = (b+cn)\left(p_{xy}-\tau|x-y|\right).
\end{aligned}
\end{equation}
Then, substituting \eqref{qxybcnpxymtauxy} into \eqref{wx}, we obtain 
\begin{equation}\label{wxequation}
w_x = \frac{b+cn}{F}\sum_{z=0}^{R-1} \left(p_{xz} - \tau|x-z|\right)^2\left(\phi_z+\lambda_z\right). 
\end{equation}
in each region $x$.

Let us define the real wage.  By \eqref{indu} and \eqref{CS}, the indirect utility of the mobile workers in each region $x$ denoted by $V_x$ is given  by
\[
V_x = \frac{a^2n}{2b}-a\sum_{z=0}^{R-1} n_z p_{zx} + \frac{b+cn}{2} \sum_{z=0}^{R-1} n_z p_{zx}^2 - \frac{c}{2}\left[\sum_{z=0}^{R-1} n_z p_{zx}\right]^2
+ w_x + \ol{q}_0.
\]
Then, it is natural to define the real wage $\omega_x$ of the mobile workers in each region $x$ by
\begin{equation}\label{realwagex}
\omega_x = w_x - a\sum_{z=0}^{R-1} n_z p_{zx} + \frac{b+cn}{2} \sum_{z=0}^{R-1} n_z p_{zx}^2 - \frac{c}{2}\left[\sum_{z=0}^{R-1} n_z p_{zx}\right]^2.
\end{equation}

Similar to \citet[Chapter 5]{FujiKrugVenab}, we adopt an ad-hoc dynamics for the migration of the population. We define the average real wage as  $\tilde{\omega}=(1/\Lambda)\sum_{z=0}^{R-1}\omega_z\lambda_z$, and assume the ad-hoc dynamics in which the population flows out of regions where the real wage is lower than the average, and flows into regions where the real wage is higher than the average.

With \eqref{nxbylamxF} and \eqref{nbyLamF}, summarizing the above equations of the prices \eqref{prices}, the nominal wage \eqref{wxequation}, the real wage \eqref{realwagex}, and the ad-hoc population dynamics, we obtain the following system:
\begin{equation}\label{2}
\left\{\begin{aligned}
&p_{xy} = \frac{a}{2(b+cn)} + \frac{c}{2(bF+cF)} \sum_{z=0}^{R-1} \lambda_z p_{zy} + \frac{\tau}{2}|x-y|,\\
&w_x(t) = \frac{b+cn}{F}\sum_{z=0}^{R-1}\left(p_{xz}(t)-\tau|x-z|\right)^2(\phi_z+\lambda_z(t)),\\
&\omega_x(t) = w_x(t) - \frac{a}{F}\sum_{z=0}^{R-1} \lambda_z(t) p_{zx}(t)\\
&\hspace{11mm} + \frac{b+cn}{2F}\sum_{z=0}^{R-1} \lambda_z(t)p_{zx}(t)^2-\frac{c}{2F^2}\left[\sum_{z=0}^{R-1} \lambda_z(t)p_{zx}(t)\right]^2,\\
&\frac{d \lambda_x}{dt}(t) = v\left[\omega_x(t) - \frac{1}{\Lambda}\sum_{z=0}^{R-1} \omega_z(t)\lambda_z(t)\right]\lambda_x(t).
\end{aligned}\right.
\end{equation}
for $(x,y)\in[0,1,\cdots,R-1]\times [0,1,\cdots,R-1]$ with an initial condition $\lambda(0)\in\mathbb{R}^{R}$, where $t\in[0, \infty)$ stands for the time variable. In the differential equation above, $v>0$ denotes the adjustment speed of the migration of the population.

In the following, we consider the system \eqref{2} with the racetrack setting. That is, $R$ regions are equidistantly distributed on a unit circle $\mathcal{C}$. For any two regions $x$ and $y$ on $\mathcal{C}$, the distance $|x-y|$ is defined by the shorter distance between them along $\mathcal{C}$. In addition, we focus on the case that the immobile population is homogeneous among regions, so
\[
\phi_x\equiv \ol{\phi} = \frac{\Phi}{R},~x=0,1,\cdots,R-1
\]
is assumed. We refer the model \eqref{2} with these settings as {\it racetrack OTT model} in the following.

\section{Stability analysis of the homogeneous stationary solution}
This section considers the homogeneous stationary solution of the racetrack OTT model, and investigates its stability for the cases of $R=2, 3$, and $4$.

\subsection{Homogeneous stationary solution}
We begin with looking for a stationary solution of \eqref{2} under the homogeneous mobile population distribution, 
\begin{equation}\label{moblstsol}
\lambda_x \equiv \ol{\lambda} = \frac{\Lambda}{R}, ~x=0,1,2,\cdots, R-1.
\end{equation}
Let us denote the prices at this state by $\ol{p}_{xy}$. Then, from \eqref{prices}
\begin{equation}\label{olpxy1}
\ol{p}_{xy} = \frac{a}{2(b+cn)} + \frac{c\ol{\lambda}}{2(bF+c\Lambda)} \sum_{z=0}^{R-1}  \ol{p}_{zy} + \frac{\tau}{2}|x-y|.
\end{equation}
Summing both sides of  \eqref{olpxy1} for $x$ yields
\begin{equation}\label{olpxy2}
\sum_{z=0}^{R-1}\ol{p}_{zy} = \frac{aRF}{2bF+c\Lambda} 
+ \frac{(bF+c\Lambda)\tau}{2bF+c\Lambda}\sum_{z=0}^{R-1}|z-y|.
\end{equation}
Substituting \eqref{olpxy2} into \eqref{olpxy1}, we obtain
\begin{equation}\label{stpTheta}
\begin{aligned}
\ol{p}_{xy} &= \frac{aF}{2bF+c\Lambda}
+ \frac{c\ol{\lambda}\tau}{2(2bF+c\Lambda)}\sum_{z=0}^{R-1} |z-y| + \frac{\tau}{2}|x-y|\\
&= \Theta + \frac{\tau}{2}|x-y|.
\end{aligned}
\end{equation}
Here, $\Theta=\Theta(\tau)$ is given by
\begin{equation}\label{defTheta}
\Theta(\tau) = \ol{\Theta}
+ \frac{c\ol{\lambda}\tau}{2(2bF+c\Lambda)}\sum_{z=0}^{R-1} |z-y|,
\end{equation}
where
\[
\ol{\Theta} = \frac{aF}{2bF+c\Lambda}.
\]
In this state, from \eqref{wxequation}, the nominal wages become 
\[
\begin{aligned}
\ol{w}_x &= \frac{b+cn}{F}\sum_{z=0}^{R-1} \left(\ol{p}_{xz}-\tau|x-z|\right)^2 \left[\ol{\phi}+\ol{\lambda}\right] \\
&= \frac{b+cn}{F}\sum_{z=0}^{R-1}\left(\Theta-\frac{\tau}{2}|x-z|\right)^2 \left[\ol{\phi}+\ol{\lambda}\right]
\end{aligned}
\]
This does not depend on the spatial variable, so it can be written as $\ol{w}_x = \ol{w}$ in any region $x$. From \eqref{realwagex}, the real wages at this state become
\[
\begin{aligned}
\ol{\omega}_x &=  \ol{w}-\frac{a\ol{\lambda}}{F}\sum_{z=0}^{R-1} \ol{p}_{zx} + \frac{(b+cn)\ol{\lambda}}{2F}\sum_{z=0}^{R-1} \ol{p}_{zx}^2 -\frac{c\ol{\lambda}}{2F^2}\left[\sum_{z=0}^{R-1} \ol{p}_{zx}\right]^2.
\end{aligned}
\]
Therefore, $\ol{\omega}_x$ is also a constant $\ol{\omega}$ in any region $x$. Thus, the ad-hoc dynamics of migration of the mobile population implies that the homogeneous population $\ol{\lambda}$ gives a stationary solution of the racetrack OTT model.

\subsection{Linearized system}
Add small perturbations $\Delta \lambda_x$, $\Delta w_x$, $\Delta p_{xy}$, and $\Delta \omega_x$ to each of the homogeneous stationary states $\ol{\lambda}$, $\ol{w}$, $\ol{p}_{xy}$, and $\ol{\omega}$. In particular, note that
\[
\sum_{x=0}^{R-1} \Delta\lambda_x \equiv 0.
\]
since the population is conserved through time. Substituting $\lambda_x = \ol{\lambda}+\Delta\lambda_x$, $w_x = \ol{w} + \Delta w_x$, $p_{xy}=\ol{p}_{xy}+\Delta p_{xy}$, and $\omega_x = \ol{\omega}+\Delta\omega_x$ for any regions $x$ and $y$ into \eqref{2}, and neglecting second and higher order terms such as $\Delta\lambda_z\Delta p_{zx}$, we obtain the linearized equations of \eqref{2}. For example, linearized price equation \eqref{prices} is given by
\[
\Delta p_{xy} = \frac{c}{2bF+c\Lambda}\sum_{z=0}^{R-1} \ol{p}_{zy}\Delta\lambda_z.
\]
As is evident from the right hand side above, $\Delta p_{xy}$ does not depend on $y$, so it can be written as 
\[\Delta p_{xy}=\Delta p_{y}.\]
Linearizing the other equations in \eqref{2} by the similar manner, we obtain the linearized system as the following:

\begin{equation}\label{linsys}
\left\{\begin{aligned}
&\Delta p_x = \frac{c}{2bF+c\Lambda}\sum_{z=0}^{R-1} \ol{p}_{zx}\Delta\lambda_z,\\
&\Delta w_x = \frac{b+cn}{F}\sum_{z=0}^{R-1} \left(\ol{p}_{xz}-\tau|x-z|\right)^2\Delta\lambda_z\\
&\hspace{10mm} + \frac{2(\ol{\phi}+\ol{\lambda})(b+cn)}{F}\sum_{z=0}^{R-1} \left(\ol{p}_{xz}-\tau|x-z|\right)\Delta p_z,\\
&\Delta\omega_x = \Delta w_x - \frac{a}{F}\sum_{z=0}^{R-1} \ol{p}_{zx}\Delta\lambda_z + \frac{b+cn}{2F}\sum_{z=0}^{R-1} \ol{p}_{zx}^2\Delta\lambda_z \\
&\hspace{20mm} - \frac{c\ol{\lambda}}{F^2}\left[\sum_{z=0}^{R-1} \ol{p}_{zx}\right]\cdot\sum_{z=0}^{R-1} \ol{p}_{zx}\Delta\lambda_z \\
&\hspace{20mm}-\frac{a\Lambda}{F}\Delta p_{x} + \frac{b\Lambda}{FR}\sum_{z=0}^{R-1} \ol{p}_{zx} \Delta p_x\\ 
&\frac{d \Delta\lambda_x}{dt}(t) = v\ol{\lambda}\Delta\omega_x.
\end{aligned}\right.
\end{equation}
We can easily confirm that 
\begin{equation}\label{sumzero}
\sum_{x=0}^{R-1} \Delta p_x 
= \sum_{x=0}^{R-1} \Delta w_x 
= \sum_{x=0}^{R-1} \Delta \omega_x
\equiv 0.
\end{equation}

\subsection{Fourier analysis}
Let$\{f_x\}_{x=0}^{R-1}$ be a sequence of data defined over $R$ points equidistantly spaced on a unit circle. The sequence $\left\{f_x\right\}_{x=0}^{R-1}$ can be represented by the Fourier series
\[
f_x = \sum_{k=0}^{R-1} \hat{f}_k e^{i2\pi kx/R},
\]
where $f_k$ is the $k$-th Fourier coefficient given by $\hat{f}_k=(1/R)\sum_{x=0}^{R-1}f_xe^{-i2\pi xk/R}$, $k=0,1,\cdots,R-1$. (See \citet[p.223]{SteinShakar} or \citet[pp.252-255]{VetKovGoy} for details.). We express the small perturbations as the Fourier series and substitute them into \eqref{linsys}. In the following let $\{\hat{\lambda}_k\}_{k=0}^{R-1}, \left\{\hat{p}_k\right\}_{k=0}^{R-1}, \left\{\hat{w}_k\right\}_{k=0}^{R-1}$, and $\left\{\hat{\omega}_k\right\}_{k=0}^{R-1}$ denote the Fourier coefficients of $\left\{\Delta \lambda_x\right\}_{x=0}^{R-1}, \left\{\Delta p_x\right\}_{x=0}^{R-1}, \left\{\Delta w_x\right\}_{x=0}^{R-1}$ and $\left\{\Delta \omega_x\right\}_{x=0}^{R-1}$, respectively. Let us see the price equation. The left hand side of the first equation of \eqref{linsys} is expressed by
\begin{equation}\label{delpFourLHS}
\sum_{k=0}^{R-1} \hat{p}_k e^{i2\pi kx/R}.
\end{equation}
On the other hand, the right hand side of the first equation of \eqref{linsys} becomes 
\begin{equation}\label{delpFourRHS}
\begin{aligned}
&\hspace{5mm}\frac{c}{2bF+c\Lambda}\sum_{z=0}^{R-1} \ol{p}_{zx}\sum_{k=0}^{R-1}\hat{\lambda}_ke^{i2\pi kz/R} \\
&= \frac{c}{2bF+c\Lambda}\sum_{k=0}^{R-1} \hat{\lambda}_k\sum_{z=0}^{R-1} \ol{p}_{zx} e^{i2\pi kz/R} \\
&= \frac{c}{2bF+c\Lambda}\sum_{k=0}^{R-1} \hat{\lambda}_k\sum_{m=-x}^{R-x-1} \ol{p}_{m} e^{i2\pi k(m+x)/R} \\
&= \frac{c}{2bF+c\Lambda}\sum_{k=0}^{R-1} \hat{\lambda}_k\left\{\sum_{m=0}^{R-1} \ol{p}_{m} e^{i2\pi km/R} \right\} e^{i2\pi kx/R},
\end{aligned}
\end{equation}
where $\ol{p}_m$ is naturally defined from \eqref{stpTheta} as 
\[
\ol{p}_m = \Theta + \frac{\tau}{2}|m|.
\]
Given that \eqref{delpFourLHS} and \eqref{delpFourRHS} are equal, 
\begin{equation}\label{phat}
\hat{p}_k = \frac{c}{2bF+c\Lambda}\left\{\sum_{m=0}^{R-1} \ol{p}_{m} e^{i2\pi km/R} \right\}\hat{\lambda}_k,~~k=1,2,\cdots, R-1
\end{equation}
holds. In the following, we do not consider $k=0$, because 
\[
\hat{\lambda}_0 = \hat{p}_0 = \hat{w}_0 = \hat{\omega}_0 \equiv 0
\]
due to \eqref{sumzero}.

By the same manner, expressing the second and the third equations of \eqref{linsys} as the Fourier series, we have
\begin{equation}\label{what}
\begin{aligned}
\hat{w}_k &= 
\frac{b+cn}{F}\left\{\sum_{m=0}^{R-1} \left(\ol{p}_{m}-\tau|m|\right)^2  e^{i2\pi km/R}\right\}\hat{\lambda}_k \\ 
&\hspace{7mm}+ \frac{2(\ol{\phi}+\ol{\lambda})(b+cn)}{F}\left\{\sum_{m=0}^{R-1}\left(\ol{p}_{m}-\tau|m|\right) e^{i2\pi km/R}\right\}\hat{p}_k,\\
&\hspace{70mm}k=1,\cdots,R-1,
\end{aligned}
\end{equation}
and 
\begin{equation}\label{omegahat}
\begin{aligned}
\hat{\omega}_k &= 
\hat{w}_k -\frac{a}{F}  \left[\sum_{m=0}^{R-1} \ol{p}_{m}  e^{i2\pi km/R}\right] \hat{\lambda}_k 
+ \frac{b+cn}{2F} \left[\sum_{m=0}^{R-1} \ol{p}_{m}^2  e^{i2\pi km/R}\right] \hat{\lambda}_k \\
&\hspace{5mm} - \frac{c\ol{\lambda}}{F^2} \left[\sum_{z=0}^{R-1} \ol{p}_{zx}\right]  \left[\sum_{m=0}^{R-1}\ol{p}_{m}  e^{i2\pi km/R}\right] \hat{\lambda}_k \\
&\hspace{5mm} + \left[-\frac{a\Lambda}{F} + \frac{b\Lambda}{FR}\sum_{z=0}^{R-1}\ol{p}_{zx} \right]\hat{p}_k,~k=1,\cdots,R-1,
\end{aligned}
\end{equation}
respectively. By \eqref{phat}, \eqref{what}, and \eqref{omegahat}, we see that 
\begin{equation}\label{Omegak}
\hat{\omega}_k = \Omega_k \hat{\lambda}_k,~k=1,2,\cdots,R-1
\end{equation}
hold, where $\Omega_k\in\mathbb{R}, ~k=1,\cdots,R-1$ are some constants.

By expressing the last differential equation of \eqref{linsys} as the Fourier series, we obtain 
\[
\begin{aligned}
\frac{d}{dt}\hat{\lambda}_k &= v\ol{\lambda}\hat{\omega}_k \\
&= v\ol{\lambda}\Omega_k \hat{\lambda}_k,~~(k=1,2,\cdots,R-1),
\end{aligned}
\]
which mean that the stability of the constant steady-state solution is determined by the sign of $\Omega_k$, i.e., the $k$-th eigenfunction is unstable if $\Omega_k>0$ and stable if $\Omega_k<0$.

\subsection{Stability and instability}
We calculate $\Omega_k,~k=1,2,\cdots,R-1$ specifically for each case of $R=2, 3,$ and $4$.

\noindent
\subsubsection{$R=2$}
Let us consider the two regional case. 
\begin{figure}[H]
\centering
\begin{tikzpicture}
 \draw(0,0) circle (1);
 \fill[black] (0,1) circle (0.1) node[above left]{$x=0$};
 \fill[black] (0,-1) circle (0.1) node[below left]{$x=1$};
\end{tikzpicture}
\caption{Equidistantly distributed two regions}
\end{figure}
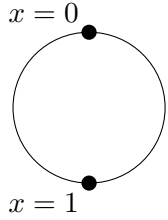

We obtain the following theorem. See Subsection \ref{proofR2} for the proof.
\begin{theo}\label{R2main}
When $R=2$, the function $\Omega_1=\Omega_1(\tau)$ satisfies $\Omega_1=0$ at only two points $\tau=0$ and $\tau=\tau^{(2)}>0$. In the interval $(0, \tau^{(2)})$, $\Omega_1>0$ holds, and in the interval $(\tau^{(2)}, \infty]$, $\Omega_1 < 0$ holds.
\end{theo}
See the sketch\footnote{The following figures \ref{fig:R2_eigen}, \ref{fig:R3_eigen}, \ref{fig:R4_eigen}, \ref{fig:R4l_eigen}, and \ref{fig:R5_eigen} show only qualitative forms of $\Omega_k$s and do not display the exact value.} in Fig. \ref{fig:R2_eigen} for the shape of the function $\Omega_1(\tau)$.

This theorem shows that the homogeneous stationary solution \eqref{moblstsol} is stable if the transport cost is sufficiently high, and unstable if the transport cost is sufficiently low when $R=2$. 

\vspace{3mm}
\noindent
{\bf Remark:}
The statement is essentially included in  \citet[Proposition 1]{OttaTabThi}. Here, however, it is discussed using the Fourier analysis, which can be easily extended to multi-regional cases.

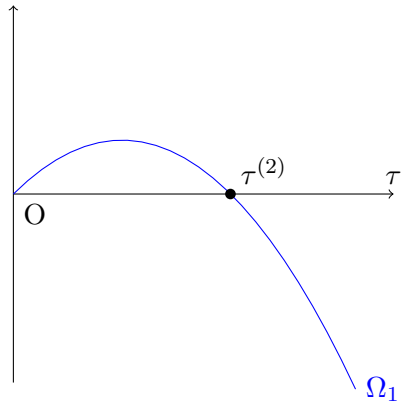
\begin{figure}[H]
\centering
\begin{tikzpicture}
 \draw[name path=xaxis,->,thin] (0,0)--(5.0,0)node[above]{$\tau$};
 \draw[->,thin] (0,-2.5)--(0,2.5)node[right]{$$};
 \draw (0,0)node[below right]{O};
 \draw[name path=Omega1,blue,thin,domain=0:4.5] plot(\x,{-0.35*pow(\x,2)+\x})node[right]{$\Omega_1$};
 \path[name intersections={of=Omega1 and xaxis}];
 \fill[black] (intersection-2) circle (0.07) node[above right]{$\tau^{(2)}$};
\end{tikzpicture}
\caption{Sketch of the graph of $\Omega_1$ when $R=2$}
\label{fig:R2_eigen}
\end{figure}

\subsubsection{$R=3$}
Let us consider the three regional case. We obtain the following theorem.
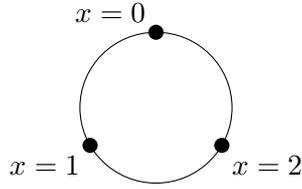
\begin{figure}[H]
\centering
\begin{tikzpicture}
 \draw(0,0) circle (1);
 \fill[black] (0,1) circle (0.1) node[above left]{$x=0$};
 \fill[black] ({-sqrt(3)/2},-0.5) circle (0.1) node[below left]{$x=1$};
 \fill[black] ({sqrt(3)/2},-0.5) circle (0.1) node[below right]{$x=2$};
\end{tikzpicture}
\caption{Equidistantly distributed three regions}
\end{figure}

\begin{theo}\label{R3main}
When $R=3$, the functions $\Omega_k=\Omega_k(\tau),~k=1,2$ satisfy $\Omega_k=0$ at only two points $\tau=0$ and $\tau=\tau^{(3)}>0$. In the interval $(0, \tau^{(3)})$, $\Omega_k>0$ hold, and in the interval $(\tau^{(3)}, \infty]$, $\Omega_k < 0$ hold for $k=1,2$.
\end{theo}

See the sketch in Fig. \ref{fig:R2_eigen} for the shape of the functions $\Omega_k(\tau),~k=1,2$.

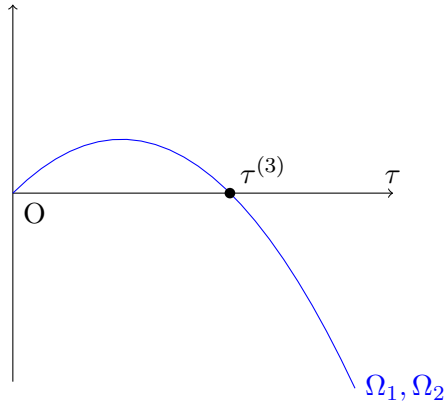
\begin{figure}[H]
\centering
\begin{tikzpicture}
 \draw[name path=xaxis,->,thin] (0,0)--(5.0,0)node[above]{$\tau$};
 \draw[->,thin] (0,-2.5)--(0,2.5)node[right]{$$};
 \draw (0,0)node[below right]{O}; 
 \draw[name path=Omega13,blue,thin,domain=0:4.5] plot(\x,{-0.35*pow(\x,2)+\x})node[right]{$\Omega_1, \Omega_2$};
 \path[name intersections={of=Omega13 and xaxis}];
 \fill[black] (intersection-2) circle (0.07) node[above right]{$\tau^{(3)}$};
\end{tikzpicture}
\caption{Sketch of the graph of $\Omega_1$ and $\Omega_2$ when $R=3$}
\label{fig:R3_eigen}
\end{figure}

This theorem, of course, shows that the homogeneous stationary solution \eqref{moblstsol} is stable if the transport cost is sufficiently high, and unstable if the transport cost is sufficiently low when $R=3$.

\subsubsection{$R=4$}
Let us consider the four regional case.
\begin{figure}[H]
\centering
\begin{tikzpicture}
 \draw(0,0) circle (1);
 \fill[black] ({-sqrt(2)/2},{sqrt(2)/2}) circle (0.1) node[above left]{$x=0$};
 \fill[black] ({-sqrt(2)/2},{-sqrt(2)/2}) circle (0.1) node[below left]{$x=1$};
 \fill[black] ({sqrt(2)/2},{-sqrt(2)/2}) circle (0.1) node[below right]{$x=2$};
 \fill[black] ({sqrt(2)/2},{sqrt(2)/2}) circle (0.1) node[above right]{$x=3$};
\end{tikzpicture}
\caption{Equidistantly distributed four regions}
\end{figure}
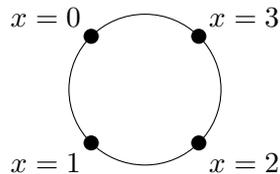

When $R=4$, the homogeneous stationary solution \eqref{moblstsol} is always unstable for any value of the transport cost. To be more precise, the following theorem holds. See Subsection \ref{proofR4} for the proof.
\begin{theo}\label{R4main}
When $R=4$, the functions $\Omega_k=\Omega_k(\tau),~k=1,3$ satisfy $\Omega_k=0$ at only two points $\tau=0$ and $\tau=\tau^{(4)}>0$. In the interval $(0, \tau^{(4)})$, $\Omega_k>0$ hold, and in the interval $(\tau^{(4)}, \infty]$, $\Omega_k < 0$ hold for $k=1,3$. Meanwhile, the function $\Omega_2=\Omega_2(\tau)$ satisfies $\Omega_2=0$ at only $\tau=0$, and $\Omega_2 > 0$ for all $\tau\in(0, \infty]$.
\end{theo}

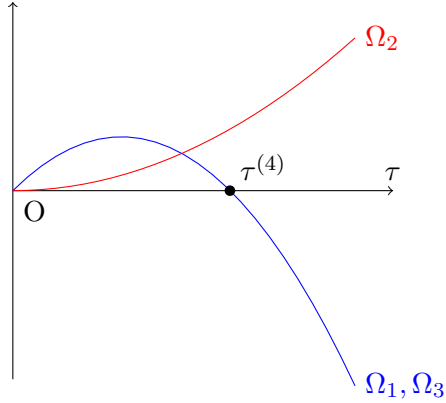
\begin{figure}[H]
\centering
\begin{tikzpicture}
 \draw[name path=xaxis,->,thin] (0,0)--(5.0,0)node[above]{$\tau$};
 \draw[->,thin] (0,-2.5)--(0,2.5)node[right]{$$};
 \draw (0,0)node[below right]{O};
 \draw[name path=Omega13,blue,thin,domain=0:4.5] plot(\x,{-0.35*pow(\x,2)+\x})node[right]{$\Omega_1, \Omega_3$};
 \draw[red,thin,domain=0:4.5] plot(\x,{0.1*pow(\x,2)})node[right]{$\Omega_2$};
 \path[name intersections={of=Omega13 and xaxis}];
 \fill[black] (intersection-2) circle (0.07) node[above right]{$\tau^{(4)}$};
\end{tikzpicture}
\caption{Sketch of the graph of $\Omega_1$, $\Omega_2$, and $\Omega_3$ when $R=4$}
\label{fig:R4_eigen}
\end{figure}

See the sketch in Fig. \ref{fig:R4_eigen} for the shape of the functions $\Omega_k(\tau),~k=1,2,3$. 

\subsection{When $R\geq 5$}
Finally, we discuss the case where $R\geq 5$. We cannot consider all $R\in \mathbb{N}$, but we obtain relatively simple results as follows, at least for the case where $R$ is a multiple of four and for the case where $R=5$.
\subsubsection{The case where $R$ is a multiple of four}
When $R\geq 5$ but is a multiple of four, the stationary solution is always unstable. To be more precise, Theorem \ref{R4lmain} below holds. See Subsection \ref{proofR4l} for the proof. 

\begin{theo}\label{R4lmain}
When $R=4l,~l=1,2,\cdots$, the function $\Omega_{2l}=\Omega_{2l}(\tau)$ satisfies $\Omega_{2l}=0$ at only $\tau=0$, and $\Omega_{2l} > 0$ for all $\tau\in(0, \infty]$.
\end{theo}

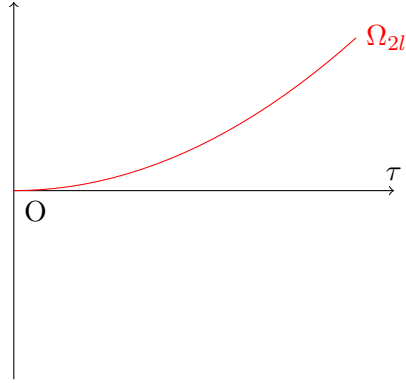
\begin{figure}[H]
\centering
\begin{tikzpicture}
 \draw[name path=xaxis,->,thin] (0,0)--(5.0,0)node[above]{$\tau$};
 \draw[->,thin] (0,-2.5)--(0,2.5)node[right]{$$};
 \draw (0,0)node[below right]{O};
 \draw[red,thin,domain=0:4.5] plot(\x,{0.1*pow(\x,2)})node[right]{$\Omega_{2l}$};
\end{tikzpicture}
\caption{Sketch of the graph of $\Omega_{2l}$ when $R=4l$}
\label{fig:R4l_eigen}
\end{figure}

See Fig. \ref{fig:R4l_eigen} for the shape of the function $\Omega_{2l}(\tau)$.

\vspace{3mm}
\noindent
{\bf Remark:} Note that this theorem does not refer to eigenfunctions with $k\neq 2l$.

\subsubsection{The case where $R=5$}
In the cases we deal with so far ($R=2,3$, and $4l,l=1,2,\cdots$), the stability of the homogeneous stationary solution does not depend on the parameter $\ol{\phi}$. However, in the case $R=5$, the appearance of the always unstable eigenfunction becomes dependent on $\ol{\phi}$. Simply stated, small values of $\ol{\phi}$ result in the emergence of the always unstable eigenfunctions, while large values of $\ol{\phi}$ allow the homogeneous stationary solution to be stabilized by the sufficiently high transport cost, as in the cases of $R=2,3$. See Subsection \ref{subsec:R5} for the proof.

\begin{theo}\label{R5main}
When $R=5$, the functions $\Omega_k(\tau)$ satisfy $\Omega_k=0$ at only two points $\tau=0$ and $\tau=\tau^{(5)}>0$ for $k=1,4$. In the interval $(0, \tau^{(5)})$, $\Omega_k>0$ hold, and in the interval $(\tau^{(5)}, \infty]$, $\Omega_k < 0$ hold for $k=1,4$. On the other hand, the behavior of the functions $\Omega_2(=\Omega_3)$ depends on the value of $\ol{\phi}\geq 0$ as the following.\vspace{2mm}\\
\noindent
\underline{When $\ol{\phi}$ is sufficiently large} \vspace{1mm} \\
The functions $\Omega_k(\tau)$ satisfy $\Omega_k=0$ at only two points $\tau=0$ and $\tau=\tau^{(5*)}>0$ for $k=2,3$. In the interval $(0, \tau^{(5*)})$, $\Omega_k>0$ hold, and in the interval $(\tau^{(5*)}, \infty]$, $\Omega_k < 0$ hold for $k=2,3$. \vspace{2mm}

\noindent
\underline{When $\ol{\phi}$ is sufficiently small} \vspace{1mm} \\
The functions $\Omega_k(\tau)$ satisfy $\Omega_k=0$ at only $\tau=0$, and $\Omega_k > 0$ for all $\tau\in(0, \infty]$ for $k=2,3$.
\end{theo}

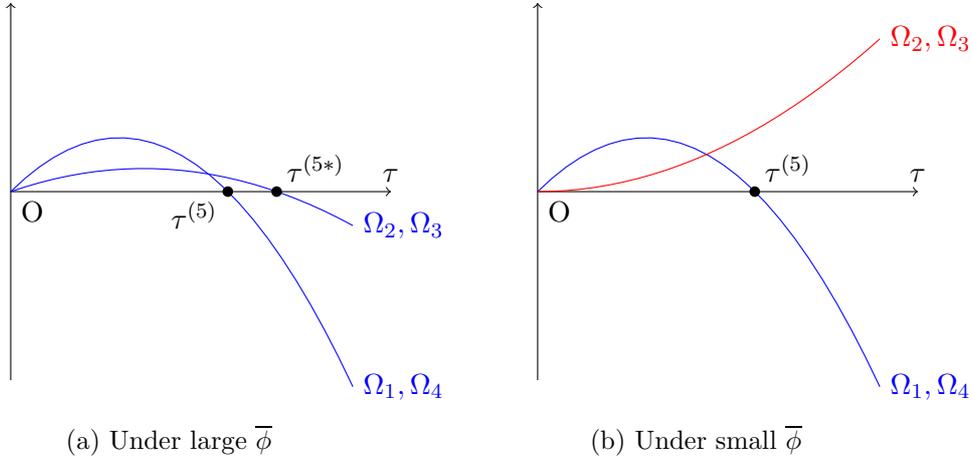
\begin{figure}[H]
 \begin{minipage}[b]{0.33\columnwidth}
  \centering
  \begin{tikzpicture}
  \draw[name path=xaxis,->,thin] (0,0)--(5.0,0)node[above]{$\tau$};
  \draw[->,thin] (0,-2.5)--(0,2.5)node[right]{$$};
  \draw (0,0)node[below right]{O};
  \draw[name path=Omega14,blue,thin,domain=0:4.5] plot(\x,{-0.35*pow(\x,2)+\x})node[right]{$\Omega_1, \Omega_4$};
  \draw[name path=Omega23,blue,thin,domain=0:4.5] plot(\x,{-0.1*pow(\x,2)+0.35*\x})node[right]{$\Omega_2, \Omega_3$};
  \path[name intersections={of=Omega14 and xaxis}];
  \fill[black] (intersection-2) circle (0.07) node[below left]{$\tau^{(5)}$};
  \path[name intersections={of=Omega23 and xaxis}];
  \fill[black] (intersection-2) circle (0.07) node[above right]{$\tau^{(5*)}$};
  \end{tikzpicture}
  \subcaption{Under large $\ol{\phi}$} \label{large_phi_eigenfunctions}
 \end{minipage}
\hspace{25mm}
 \begin{minipage}[b]{0.33\columnwidth}
  \centering
  \begin{tikzpicture}
  \draw[name path=xaxis,->,thin] (0,0)--(5.0,0)node[above]{$\tau$};
  \draw[->,thin] (0,-2.5)--(0,2.5)node[right]{$$};
  \draw (0,0)node[below right]{O};
  \draw[name path=Omega13,blue,thin,domain=0:4.5] plot(\x,{-0.35*pow(\x,2)+\x})node[right]{$\Omega_1, \Omega_4$};
  \draw[red,thin,domain=0:4.5] plot(\x,{0.1*pow(\x,2)})node[right]{$\Omega_2, \Omega_3$};
  \path[name intersections={of=Omega13 and xaxis}];
  \fill[black] (intersection-2) circle (0.07) node[above right]{$\tau^{(5)}$};
  \end{tikzpicture}
  \subcaption{Under small $\ol{\phi}$}\label{small_phi_eigenfunctions}
 \end{minipage}
 \caption{Sketch of the graph of $\Omega_1$, $\Omega_2$, $\Omega_3$, and $\Omega_4$ when $R=5$}
\label{fig:R5_eigen}
\end{figure}

\vspace{3mm}
See Fig.\ref{fig:R5_eigen} for the shape of the functions $\Omega_k(\tau)$, $k=1,2,3,4$.

\vspace{3mm}
\noindent
{\bf Remark}

\noindent
Note that the value of $\tau^{(5)}$ varies with the value of $\ol{\phi}$ although $\tau^{(5)}$ in Fig.\ref{fig:R5_eigen}\subref{large_phi_eigenfunctions} and that in Fig.\ref{fig:R5_eigen}\subref{small_phi_eigenfunctions} are depicted as if they were the same value.

\section{Discussion and conclusion}
We extend the model by \citet{OttaTabThi} to a multi-regional economy and investigate the stability of the homogeneous stationary solution. The results show that up to three regions, the stationary solution is stable when the transport cost is high and unstable when the cost is low, which is a common property of the standard models of new economic geography such as the core-periphery model. However, when the number of regions reaches to a multiple of four, an eigenfunction which is unstable under any values of the transport cost appears, and thus the homogeneous stationary solution is no longer stable. 

A hypothetical explanation of the mechanism by which the homogeneous stationary solution is always unstable when the number of regions is a multiple of four, although it is difficult to state it intuitively, may be as follows. Basically, each region is considered to be in competition with each other for population. Regions located near larger regions would be likely to have their population absorbed by the larger regions. On the other hand, regions that are far enough away from large regions may likely to keep their population. When the number of regions is a multiple of four, for every region, there is a ``sister region" so to speak, located exactly opposite to it on the racetrack. Such sister regions have the least growth-inhibiting effect on each other due to their greatest distance. Therefore, there would be a potential for destabilization in the form of pairs of such sister regions growing without interfering with each other\footnote{In this case, regions adjacent to the growing sister regions may reduce their population.}. Indeed, the fact that the frequency of the destabilizing eigenfunction is $2l~(l=1,2,3,\cdots)$ when the number of regions is $4l$ seems to support this.

When $R=4l$, the calculation of eigenvalue for the $2l$-th eigenfunction is greatly simplified. Especially, the effect of the Fourier coefficient $\hat{p}_{2l}$ being zero as in \eqref{phat2l} is particularly significant. In other cases of $R\neq 4l$, the coefficient  $\hat{p}_k$ expressed by \eqref{phat} is not become $0$ in general, and thus the expression for the $k$-th eigenvalue is so complex that it becomes more and more difficult to determine its sign as the number of regions increases. If there is some regular relationship between the number of regions and the eigenvalues, it would be possible to explore the stability of the homogeneous stationary solution for the general cases of $R\geq 6$.

The result of this paper suggests the theoretical possibility that the number of regions itself can be a cause of economic agglomeration. 
The property that the homogeneous stationary solution is always unstable when the number of regions is a multiple of four does not hold for the standard multi-regional core-periphery model on the racetrack. In fact, \citet{IkeAkaKon} have shown in the context of bifurcation analysis that when the number of regions is even, the homogeneous stationary solution of the racetrack core-periphery model is stable for a sufficiently high transport cost. Furthermore, our result suggests that the characteristics of the OTT model, when compared to the standard core-periphery model, become more pronounced when considering its multi-regional version. It also may provide a criterion to consider when modeling a multi-regional spatial economy as to whether it is more appropriate to rely on the core-periphery model or the OTT model (depending, of course, on the spatial economy itself being modeled).

\section{Appendix}

\subsection{Proof for Theorem \ref{R2main}}\label{proofR2}

The distances between two evenly distributed regions are
\begin{equation}\label{disR2}
|x-y|=
\left\{
\begin{array}{l}
|0|=0,~\mbox{if~}x= y,\\
|1|=\pi,~\mbox{if~}x\neq y.
\end{array}
\right.
\end{equation}
for $x,y\in\left\{0, 1\right\}$. From \eqref{stpTheta} and \eqref{disR2}, we see that
\begin{equation}\label{stp0p1R2}
\left\{
\begin{aligned}
&\ol{p}_0 = \Theta + \frac{\tau}{2}|0| = \Theta, \\
&\ol{p}_1 = \Theta + \frac{\tau}{2}|1| = \Theta + \frac{\pi}{2}\tau.
\end{aligned}
\right.
\end{equation}
We only have to discuss the effective frequency $k=1$. From \eqref{phat} and \eqref{stp0p1R2}, we have
\begin{equation}\label{p1hR2}
\begin{aligned}
\hat{p}_1 &= 
 \frac{c}{2bF+c\Lambda}\left\{\sum_{m=0}^{1} \ol{p}_{m} e^{i\pi m} \right\}\hat{\lambda}_1 \\
&= \frac{c}{2bF+c\Lambda}\left\{\ol{p}_0 - \ol{p}_1\right\}\hat{\lambda}_1\\
&= -\frac{c\pi\tau}{2(2bF+c\Lambda)} \hat{\lambda}_1.
\end{aligned}
\end{equation}
Since 
\[
\sum_{x=0}^1 |x-y| = 0+\pi=\pi,
\]
we see from \eqref{defTheta} that
\begin{equation}\label{ThThbR2}
\Theta(\tau) = \ol{\Theta} + \frac{c\ol{\lambda}\pi}{2(2bF+c\Lambda)}\tau.
\end{equation}
Then, from \eqref{what}, \eqref{stp0p1R2}, \eqref{p1hR2}, and \eqref{ThThbR2}, we have
\begin{equation}\label{w1hR2}
\begin{aligned}
\hat{w}_1 
&= \left[\frac{b+cn}{F}\left\{\ol{\Theta}\pi\tau+\frac{c\ol{\lambda}\pi^2\tau^2}{2(2bF+c\Lambda)}-\frac{\pi^2\tau^2}{4}\right\}\right.\\
&\left.\hspace{35mm}-\frac{2(\ol{\phi}+\ol{\lambda})(b+cn)c\pi^2\tau^2}{4F(2bF+c\Lambda)}\right]\hat{\lambda}_1
\end{aligned}
\end{equation}
By the same manner, from \eqref{omegahat}, \eqref{stp0p1R2}, \eqref{p1hR2}, and \eqref{ThThbR2}. we have
\begin{equation}\label{omegahatR2k1}
\begin{aligned}
\hat{\omega}_1 
&=
\hat{w_1} + \frac{a\pi\tau}{2F}\hat{\lambda_1}
-\frac{b+cn}{2F}\left[\pi\ol{\Theta}\tau+\frac{c\ol{\lambda}\pi^2\tau^2}{2(2bF+c\Lambda)}+\frac{\pi^2\tau^2}{4}\right] \hat{\lambda_1}\\
&\hspace{10mm}+\frac{c\ol{\lambda}}{F^2}\left[\pi\ol{\Theta}\tau+\frac{c\ol{\lambda}\pi^2\tau^2}{2(2bF+c\Lambda)}+\frac{\pi^2\tau^2}{4}\right]\hat{\lambda_1}\\
&\hspace{10mm}+\left[\frac{a\Lambda c\pi\tau}{2F(2bF+c\Lambda)}-\frac{b\Lambda}{2F}
\left(
\frac{\ol{\Theta}c\pi\tau}{2F(2bF+c\Lambda)}\right.\right.\\
&\left.\left.\hspace{30mm}+\frac{c^2\ol{\lambda}\pi^2\tau^2}{2(2bF+c\Lambda)^2}+\frac{c\pi^2\tau^2}{4(2bF+c\Lambda)}
\right)
\right]\hat{\lambda}_1
\end{aligned}
\end{equation}
Combining \eqref{w1hR2} and \eqref{omegahatR2k1}, we get
\[
\Omega_1
= A^{(2)}\tau^2 + B^{(2)}\tau,
\]
where
\[
\left\{
\begin{aligned}
&A^{(2)} = 
\frac{b+cn}{F}\left(\frac{c\ol{\lambda}\pi^2}{2(2bF+c\Lambda)}-\frac{\pi^2}{4}\right)
- \frac{2(\ol{\phi}+\ol{\lambda})(b+cn)}{F}\cdot \frac{c}{2bF+c\Lambda}\cdot\frac{\pi^2}{4}\\
&\hspace{10mm}- \frac{b+cn}{2F}\left(\frac{c\ol{\lambda}\pi^2}{2(2bF+c\Lambda)} +  \frac{\pi^2}{4}\right) + \frac{c\ol{\lambda}}{F^2}\left(\frac{c\ol{\lambda}\pi^2}{2(2bF+c\Lambda)} + \frac{\pi^2}{4}\right) \\
&\hspace{10mm}-\left[\frac{b\Lambda c^2\ol{\lambda}\pi^2}{4F(2bF+c\Lambda)^2}+\frac{b\Lambda c\pi^2}{8F(2bF+c\Lambda)}\right]  < 0,\\
&B^{(2)} = \frac{b+cn}{F}\pi\ol{\Theta} + \frac{a\pi}{2F} - \frac{b+cn}{2F}\pi\ol{\Theta} + \frac{c\ol{\lambda}}{F^2}\pi\ol{\Theta} \\
&\hspace{10mm} + \frac{a\Lambda c\pi}{2F(2bF+c\Lambda)} - \frac{b\Lambda\ol{\Theta}c\pi}{2F(2bF+c\Lambda)} > 0.
\end{aligned}
\right.
\]
We can show by careful calculation that $A^{(2)}<0$ and $B^{(2)}>0$. These facts show that $\Omega_1$ is a quadratic function passing through the origin with respect to $\tau\geq 0$, and that $\Omega_1<0$ for sufficiently large $\tau$ and $\Omega_1>0$ for sufficiently small $\tau$. In addition, there is only one $\tau>0$ for which $\Omega_1=0$. \qed

\subsection{Proof for Theorem \ref{R3main}}

The distances between three evenly distributed regions are
\begin{equation}\label{disR3}
|x-y|=
\left\{
\begin{aligned}
&|0|=0,~\mbox{if~}x= y,\\
&|1|=\frac{2\pi}{3},~\mbox{if~}x=y+1\pmod 3,\\
&|2|=\frac{2\pi}{3},~\mbox{if~}x=y+2\pmod 3.
\end{aligned}
\right.
\end{equation}
for $x,y\in\left\{0,1,2\right\}$. From \eqref{stpTheta} and \eqref{disR3}, we see that
\begin{equation}\label{stp0p1R3}
\left\{
\begin{aligned}
&\ol{p}_0 = \Theta + \frac{\tau}{2}|0| = \Theta, \\
&\ol{p}_1 = \Theta + \frac{\tau}{2}|1| = \Theta + \frac{\pi}{3}\tau, \\
&\ol{p}_2 = \Theta + \frac{\tau}{2}|2| = \Theta + \frac{\pi}{3}\tau.
\end{aligned}
\right.
\end{equation}
We only need to discuss two effective frequencies $k=1$ and $2$. From \eqref{phat} and \eqref{stp0p1R3}, we have
\begin{equation}\label{p1hR3}
\begin{aligned}
\hat{p}_1 &= 
 \frac{c}{2bF+c\Lambda}\left\{\sum_{m=0}^{2} \ol{p}_{m} e^{i2\pi m/3} \right\}\hat{\lambda}_1 \\
&= \frac{c}{2bF+c\Lambda}\left\{\ol{p}_0 + \ol{p}_1\left(e^{i\frac{2\pi}{3}} + e^{i\frac{4\pi}{3}}\right)\right\}\hat{\lambda}_1\\
&= - \frac{c\pi\tau}{3(2bF+c\Lambda)} \hat{\lambda}_1.
\end{aligned}
\end{equation}
Similarly,
\begin{equation}\label{p2hR3}
\hat{p}_2 = - \frac{c\pi\tau}{3(2bF+c\Lambda)} \hat{\lambda}_2.
\end{equation}
Since 
\[
\sum_{x=0}^2 |x-y| = 0+\frac{2\pi}{3} + \frac{2\pi}{3}=\frac{4\pi}{3},
\]
we see from \eqref{defTheta} that
\begin{equation}\label{ThThbR3}
\Theta(\tau) = \ol{\Theta} + \frac{c\Lambda2\pi\tau}{9(2bF+c\Lambda)} .
\end{equation}
Then, from \eqref{what}, \eqref{stp0p1R3}, \eqref{p1hR3}, \eqref{p2hR3}, and \eqref{ThThbR3}, we have
\begin{equation}\label{w1hR3}
\left\{
\begin{aligned}
&\hat{w}_1 = \left[
\frac{b+cn}{F}\left\{\frac{2\pi}{3}\ol{\Theta}\tau + \frac{c\Lambda 4\pi^2}{27(2bF+c\Lambda)}\tau^2 - \frac{\pi^2}{9}\tau^2\right\}\right.\\
&\hspace{20mm}\left. - \frac{2(\ol{\phi}+\ol{\lambda})(b+cn)}{F}\cdot\frac{c\pi^2}{9(2bF+c\Lambda)}\tau^2
\right]\hat{\lambda}_1, \\
&\hat{w}_2 = \left[
\frac{b+cn}{F}\left\{\frac{2\pi}{3}\ol{\Theta}\tau + \frac{c\Lambda 4\pi^2}{27(2bF+c\Lambda)}\tau^2 - \frac{\pi^2}{9}\tau^2\right\}\right.\\
&\hspace{20mm}\left. - \frac{2(\ol{\phi}+\ol{\lambda})(b+cn)}{F}\cdot\frac{c\pi^2}{9(2bF+c\Lambda)}\tau^2
\right]\hat{\lambda}_2.
\end{aligned}
\right.
\end{equation}
By the same manner, from \eqref{omegahat}, \eqref{stp0p1R3}, \eqref{p1hR3}, \eqref{p2hR3}, and \eqref{ThThbR3}, we have
\begin{equation}\label{omegahat12R3}
\left\{
\begin{aligned}
&\hat{\omega}_1 = 
\hat{w}_1 + \frac{a}{F}\left[\frac{\pi\tau}{3}\right]\hat{\lambda}_1\\
&\hspace{10mm} - \frac{b+cn}{2F}\left[\frac{2\pi}{3}\left(\ol{\Theta}+\frac{c\Lambda2\pi\tau}{9(2bF+c\Lambda)}\right)\tau + \frac{\pi^2\tau^2}{9}\right]\hat{\lambda}_1\\
&\hspace{10mm} + \frac{c\ol{\lambda}}{F^2}\left[3\left(\ol{\Theta}+\frac{c\Lambda2\pi\tau}{9(2bF+c\Lambda)}\right)+\frac{2\pi}{3}\tau\right]\left[\frac{\pi\tau}{3}\right]\hat{\lambda}_1 \\
&\hspace{10mm} + \left[\frac{a\Lambda c\pi}{3F(2bF+c\Lambda)}\tau - \frac{b\Lambda}{3F}\left(\frac{\ol{\Theta}c\pi}{2bF+c\Lambda}\tau \right.\right.\\
&\hspace{25mm}\left.\left. + \frac{c^2\Lambda 2\pi^2\tau^2}{9(2bF+c\Lambda)^2} + \frac{2\pi^2c\tau^2}{9(2bF+c\Lambda)}\right)\right]\hat{\lambda}_1,\\
&\hat{\omega}_2 = 
\hat{w}_2 + \frac{a}{F}\left[\frac{\pi\tau}{3}\right]\hat{\lambda}_2\\
&\hspace{10mm} - \frac{b+cn}{2F}\left[\frac{2\pi}{3}\left(\ol{\Theta}+\frac{c\Lambda2\pi\tau}{9(2bF+c\Lambda)}\right)\tau + \frac{\pi^2\tau^2}{9}\right]\hat{\lambda}_2\\
&\hspace{10mm} + \frac{c\ol{\lambda}}{F^2}\left[3\left(\ol{\Theta}+\frac{c\Lambda2\pi\tau}{9(2bF+c\Lambda)}\right)+\frac{2\pi}{3}\tau\right]\left[\frac{\pi\tau}{3}\right]\hat{\lambda}_2 \\
&\hspace{10mm} + \left[\frac{a\Lambda c\pi}{3F(2bF+c\Lambda)}\tau - \frac{b\Lambda}{3F}\left(\frac{\ol{\Theta}c\pi}{2bF+c\Lambda}\tau\right.\right.\\
&\left.\left.\hspace{25mm} + \frac{c^2\Lambda 2\pi^2\tau^2}{9(2bF+c\Lambda)^2} + \frac{2\pi^2c\tau^2}{9(2bF+c\Lambda)}\right)\right]\hat{\lambda}_2.
\end{aligned}
\right.
\end{equation}
Combining \eqref{w1hR3} and \eqref{omegahat12R3}, we get
\[
\Omega_1=\Omega_2
= A^{(3)}\tau^2 + B^{(3)}\tau,
\]
where
\[
\left\{
\begin{aligned}
&A^{(3)} = 
\frac{b+cn}{F}\cdot \frac{c\Lambda 4\pi^2}{27(2bF+c\Lambda)} - \frac{b+cn}{F}\cdot \frac{\pi^2}{9} - \frac{2(\ol{\phi}+\ol{\lambda})(b+cn)}{F}\cdot\frac{c\pi^2}{9(2bF+c\Lambda)}\\
&\hspace{10mm}- \frac{b+cn}{2F}\cdot\frac{c\Lambda4\pi^2}{9(2bF+c\Lambda)} - \frac{b+cn}{2F}\cdot\frac{\pi^2}{9}\\
&\hspace{10mm}+ \frac{c\ol{\lambda}}{F^2}\cdot\frac{c\Lambda 2\pi^2}{9(2bF+c\Lambda)} + \frac{c\ol{\lambda}}{F^2}\cdot\frac{2\pi^2}{9}\\
&\hspace{10mm}- \frac{b\Lambda}{F}\left(\frac{c^2\Lambda 2\pi^2}{9(2bF+c\Lambda)^2} + \frac{2\pi^2c}{9(2bF+c\Lambda)}\right) <0,\\
&B^{(3)} = \frac{b+cn}{F}\cdot\frac{2\pi}{3}\ol{\Theta} \\
&\hspace{15mm} - \frac{b+cn}{2F}\frac{2\pi}{3}\ol{\Theta} + \frac{c\ol{\lambda}}{F^2}\pi\ol{\Theta} \\
&\hspace{15mm} + \frac{a\Lambda c\pi}{3F(2bF+c\Lambda)} - \frac{b\Lambda}{3F}\frac{\ol{\Theta}c\pi}{2bF+c\Lambda} > 0.
\end{aligned}\right.
\]
We can show by careful calculation that $A^{(3)}<0$ and $B^{(3)}>0$.  These facts show that $\Omega_1(=\Omega_2)$ is a quadratic function passing through the origin with respect to $\tau\geq 0$, and that $\Omega_1(=\Omega_2)<0$ for sufficiently large $\tau$ and $\Omega_1(=\Omega_2)>0$ for sufficiently small $\tau$. In addition, there is only one $\tau>0$ for which $\Omega_1(=\Omega_2)=0$. \qed

\subsection{Proof for Theorem \ref{R4main}} \label{proofR4}

The distances between four evenly distributed regions are
\begin{equation}\label{disR4}
|x-y|=
\left\{
\begin{aligned}
&|0|=0,~\hspace{1mm}\mbox{if~}x= y,\\
&|1|=\frac{\pi}{2},~\mbox{if~}x=y+1\pmod 4,\\
&|2|=\pi,\hspace{1mm}~\mbox{if~}x=y+2\pmod 4,\\
&|3|=\frac{\pi}{2},~\mbox{if~}x=y+3\pmod 4
\end{aligned}
\right.
\end{equation}
for $x,y\in\left\{0,1,2,3\right\}$. From \eqref{stpTheta} and \eqref{disR4}, we see that
\begin{equation}\label{stp0p1R4}
\left\{
\begin{aligned}
&\ol{p}_0 = \Theta + \frac{\tau}{2}|0| = \Theta, \\
&\ol{p}_1 = \Theta + \frac{\tau}{2}|1| = \Theta + \frac{\pi}{4}\tau, \\
&\ol{p}_2 = \Theta + \frac{\tau}{2}|2| = \Theta + \frac{\pi}{2}\tau, \\
&\ol{p}_3 = \Theta + \frac{\tau}{2}|3| = \Theta + \frac{\pi}{4}\tau.
\end{aligned}
\right.
\end{equation}
We only need to discuss three effective frequencies $k=1$, $2$, and $3$. From \eqref{phat} and \eqref{stp0p1R4}, we have
\begin{equation}\label{phR4}
\left\{
\begin{aligned}
&\hat{p}_1 = - \frac{c}{2bF+c\Lambda}\left(\frac{\pi\tau}{2} \right)\hat{\lambda}_1,\\
&\hat{p}_2 = 0,\\
&\hat{p}_3 = -\frac{c}{2bF+c\Lambda}\left(\frac{\pi\tau}{2}\right)\hat{\lambda}_3. 
\end{aligned}
\right.
\end{equation}
Since 
\[
\sum_{x=0}^3 |x-y| = 0+\frac{\pi}{2} + \pi + \frac{\pi}{2} = 2\pi,
\]
we see from \eqref{defTheta} that
\begin{equation}\label{ThThbR4}
\Theta(\tau) = \ol{\Theta} + \frac{c\ol{\lambda}\pi}{2bF+c\Lambda}\cdot \tau.
\end{equation}
Then, from \eqref{what}, \eqref{stp0p1R4}, \eqref{phR4}, and \eqref{ThThbR4}, we have
\begin{equation}\label{whR4}
\left\{
\begin{aligned}
&\hat{w}_1 = \left[
\frac{b+cn}{F}\left\{\pi\ol{\Theta}\tau + \frac{c\ol{\lambda}\pi^2}{2bF+c\Lambda}\cdot \tau^2 - \frac{\pi^2}{4}\tau^2\right\}\right.\\
&\hspace{35mm}\left. - \frac{2(\ol{\phi}+\ol{\lambda})(b+cn)c\pi^2}{4F(2bF+c\Lambda)}
\cdot\tau^2
\right]\hat{\lambda}_1, \\
&\hat{w}_2 = \frac{b+cn}{F}\left\{\frac{\pi^2\tau^2}{8}\right\}\hat{\lambda}_2, \\
&\hat{w}_3 = \left[
\frac{b+cn}{F}\left\{\pi\ol{\Theta}\tau + \frac{c\ol{\lambda}\pi^2}{2bF+c\Lambda}\cdot \tau^2 - \frac{\pi^2}{4}\tau^2\right\}\right. \\
&\hspace{35mm}\left. - \frac{2(\ol{\phi}+\ol{\lambda})(b+cn)c\pi^2}{4F(2bF+c\Lambda)}
\cdot\tau^2
\right]\hat{\lambda}_3.
\end{aligned}
\right.
\end{equation}
By the same manner, from \eqref{omegahat}, \eqref{stp0p1R4},  \eqref{phR4}, and \eqref{ThThbR4}, we have
\begin{equation}\label{omegahat123R4}
\left\{
\begin{aligned}
&\hat{\omega}_1 = 
\hat{w}_1 
+ \frac{a\pi}{2F}\tau\hat{\lambda}_1 
- \frac{b+cn}{2F}\left[\pi\ol{\Theta}\tau + \frac{c\ol{\lambda}\pi^2}{2bF+c\Lambda}\tau^2 + \frac{\pi^2\tau^2}{4}\right]\hat{\lambda}_1 \\
&\hspace{10mm} + \frac{c\ol{\lambda}}{F^2}\left(2\pi\ol{\Theta}\tau + \frac{2c\ol{\lambda}\pi^2}{2bF+c\Lambda}\tau^2+\frac{\pi^2\tau^2}{2}\right)\hat{\lambda}_1 \\
&\hspace{10mm} + \left[\frac{a\Lambda}{F}\cdot\frac{c\pi\tau}{2(2bF+c\Lambda)} - \frac{b\Lambda}{F}\ol{\Theta}\frac{c\pi\tau}{2(2bF+c\Lambda)} \right. \\
&\hspace{20mm} \left. - \frac{b\Lambda}{F}\cdot\frac{c^2\ol{\lambda}\pi^2\tau^2}{2(2bF+c\Lambda)^2} - \frac{b\Lambda}{4F}\cdot\frac{c\pi^2\tau^2}{2(2bF+c\Lambda)}\right] \hat{\lambda}_1, \\
&\hat{\omega}_2 = 
\hat{w}_2 + \frac{b+cn}{2F}\cdot\frac{\pi^2\tau^2}{8}\cdot\hat{\lambda}_2, \\
&\hat{\omega}_3 =
\hat{w}_3 
+ \frac{a\pi}{2F}\tau\hat{\lambda}_3 
- \frac{b+cn}{2F}\left[\pi\ol{\Theta}\tau + \frac{c\ol{\lambda}\pi^2}{2bF+c\Lambda}\tau^2 + \frac{\pi^2\tau^2}{4}\right]\hat{\lambda}_3 \\
&\hspace{10mm} + \frac{c\ol{\lambda}}{F^2}\left(2\pi\ol{\Theta}\tau + \frac{2c\ol{\lambda}\pi^2}{2bF+c\Lambda}\tau^2+\frac{\pi^2\tau^2}{2}\right)\hat{\lambda}_3 \\
&\hspace{10mm} + \left[\frac{a\Lambda}{F}\cdot\frac{c\pi\tau}{2(2bF+c\Lambda)} - \frac{b\Lambda}{F}\ol{\Theta}\frac{c\pi\tau}{2(2bF+c\Lambda)}\right. \\
&\hspace{20mm} \left.  - \frac{b\Lambda}{F}\cdot\frac{c^2\ol{\lambda}\pi^2\tau^2}{2(2bF+c\Lambda)^2} - \frac{b\Lambda}{4F}\cdot\frac{c\pi^2\tau^2}{2(2bF+c\Lambda)}\right] \hat{\lambda}_3.
\end{aligned}
\right.
\end{equation}
Combining \eqref{whR4} and \eqref{omegahat123R4}, we get
\[
\begin{aligned}
&\Omega_1(=\Omega_3)
= A^{(4)}\tau^2 + B^{(4)}\tau, \\
&\Omega_2 = C^{(4)}\tau^2,
\end{aligned}
\]
where
\[
\left\{\begin{aligned}
&A^{(4)} = \frac{b+cn}{F}\left(\frac{c\ol{\lambda}\pi^2}{2bF+c\Lambda}-\frac{\pi^2}{4}\right) - \frac{2(\ol{\phi}+\ol{\lambda})(b+cn)c\pi^2}{4F(2bF+c\Lambda)} \\
&\hspace{20mm} - \frac{b+cn}{2F}\left(\frac{c\ol{\lambda}\pi^2}{2bF+c\Lambda}+\frac{\pi^2}{4}\right) \\
&\hspace{20mm} + \frac{c\ol{\lambda}}{F^2}\left(\frac{2c\ol{\lambda}\pi^2}{2bF+c\Lambda} + \frac{\pi^2}{2}\right) \\
&\hspace{20mm} -\frac{b\Lambda}{F}\frac{c^2\ol{\lambda}\pi^2}{2(2bF+c\Lambda)^2}
- \frac{b\Lambda}{4F}\frac{c\pi^2}{2(2bF+c\Lambda)}<0,\\
&B^{(4)} = 
\frac{b+cn}{F}\pi\ol{\Theta}
+ \frac{a\pi}{2F} - \frac{b+cn}{2F}\pi\ol{\Theta}
+ \frac{c\ol{\lambda}}{F^2}2\ol{\Theta}\pi \\
&\hspace{19mm} + \frac{a\Lambda}{F}\cdot\frac{c\pi}{2(2bF+c\Lambda)} - \frac{b\Lambda}{F}\ol{\Theta}\frac{c\pi}{2(2bF+c\Lambda)}>0,\\
&C^{(4)} = \frac{(b+cn)\pi^2}{8F} + \frac{b+cn}{2F}\cdot\frac{\pi^2}{8}>0.\hspace{25mm}
\end{aligned}\right.
\]
We can show by careful calculation that $A^{(4)}<0$, $B^{(4)}>0$, and $C^{(4)}>0$. These facts show that $\Omega_1(=\Omega_3)$ is a quadratic function passing through the origin with respect to $\tau\geq 0$, and that $\Omega_1(=\Omega_3)<0$ for sufficiently large $\tau$ and $\Omega_1(=\Omega_3)>0$ for sufficiently small $\tau$. In addition, there is only one $\tau>0$ for which $\Omega_1(=\Omega_3)=0$. The crucial difference from the case of $R=2$ and $3$ is that here $\Omega_2$ is positive for all $\tau>0$. \qed

\subsection{Proof for Theorem \ref{R4lmain}} \label{proofR4l}
The distances between $R(=4l)$ evenly distributed regions are
\begin{equation}\label{disR4l}
|x-y|=
\left\{
\begin{aligned}
&|0|=0,~\hspace{12.5mm}\mbox{if~}x= y,\\
&|1|=\frac{2\pi}{4l},\hspace{9.5mm}~\mbox{if~}x=y+1\pmod {4l},\\
&|2|=\frac{2\pi}{4l}\times 2,\hspace{3mm}~\mbox{if~}x=y+2\pmod {4l},\\
&\hspace{10mm}\vdots\\
&|m|=\frac{2\pi}{4l}\times m,~\mbox{if~}x=y+m\pmod {4l}
\end{aligned}
\right.
\end{equation}
for $x,y\in\left\{0,1,2,\cdots,4l\right\}$. 

For $k=2l$, the term $\sum_m \ol{p}_{m} e^{i2\pi km/R}$ in \eqref{phat} becomes 
\[
\begin{aligned}
\sum_{m=0}^{R-1} \ol{p}_m e^{i\pi m}
&= \ol{p}_0 e^0 + \ol{p}_1 e^{i\pi} + \ol{p}_2 e^{i2\pi} + \ol{p}_3 e^{i3\pi} + \cdots +\ol{p}_{2l} e^{i2l\pi} + \cdots + \ol{p}_{4l-1} e^{i(4l-1)\pi}\\
&= \ol{p}_0 -\ol{p}_1+\ol{p}_2-\ol{p}_3+\cdots+\ol{p}_{2l}-\cdots-\ol{p}_{4l-1}\\
&= \ol{p}_0 -2\ol{p}_{1} +2\ol{p}_2 -2\ol{p}_3 +\cdots -2\ol{p}_{2l-1}+\ol{p}_{2l} \\
&= -2\left(|1|+|3|+|5|+\cdots+|2l-1|\right)\\
&\hspace{10mm} +2\left(|2|+|4|+\cdots+|2l-2|\right) + |2l|\\
&= -2\cdot\frac{2\pi}{4l}(1+3+5+\cdots+(2l-1)) \\
&\hspace{10mm} +2\cdot\frac{2\pi}{4l}(2+4+\cdots+(2l-2)) + \frac{2\pi}{4l}\cdot 2l \\
&= -\frac{\pi}{l}\cdot l^2 + \frac{\pi}{l}\cdot l(l-1) + \pi \\
&= 0.
\end{aligned}
\]
Here we use the fact that $1+3+\cdots+2l-1=l^2$ and $2+4+\cdots+2l-2=l(l-1)$. Therefore, we have
\begin{equation}\label{phat2l}
\begin{aligned}
\hat{p}_{2l} &= \frac{c}{2bF+c\Lambda}\sum_{m=0}^{R-1} \ol{p}_m e^{i\pi m} \\
&= 0.
\end{aligned}
\end{equation}
For $k=2l$, the term $\sum_{m=0}^{R-1}(\ol{p}_m-\tau|m|)^2 e^{i2\pi k/R}$ in \eqref{what} becomes
\[
\begin{aligned}
\sum_{m=0}^{R-1} (\ol{p}_m-\tau|m|)^2 e^{i\pi m} 
&= \sum_{m=0}^{R-1} \left(\Theta-\frac{\tau}{2}|m|\right)^2 e^{i\pi m}\\
&= \frac{\tau^2\pi^2}{8l}
\end{aligned}
\]
Here we use the fact that $1^2+3^2+\cdots+(2l-1)^2 = \frac{2}{3}l(l+1)(2l+1)-2l(l+1)+l$ and $2^2+4^2+\cdots(2l-2)^2 = \frac{2}{3}l(l+1)(2l+1)-4l^2$. Therefore, we have 
\begin{equation}\label{what2l}
\begin{aligned}
\hat{w}_{2l}&= \frac{b+cn}{F} \left\{\sum_{m=0}^{R-1} (\ol{p}_m-\tau|m|)^2e^{i\pi m}\right\} \hat{\lambda}_{2l} \\
&=\frac{b+cn}{F}\cdot\frac{\tau^2\pi^2}{8l} \hat{\lambda}_{2l}.
\end{aligned}
\end{equation}

Finally, for $k=2l$ the term $\sum_m \ol{p}_m^2e^{i2\pi km/R}$ in \eqref{omegahat} becomes
\[
\begin{aligned}
\sum_{m=0}^{R-1}\ol{p}_m^2e^{i\pi m}
&= \sum_{m=0}^{R-1} \left(\Theta+\frac{\tau}{2}|m|\right)^2e^{i\pi m}\\
&= \sum_{m=0}^{R-1} \left(\Theta^2+\tau\Theta|m|+\frac{\tau^2}{4}|m|^2\right)e^{i\pi m}\\
&= \frac{\tau^2\pi^2}{8l}.
\end{aligned}
\]
by the same calculation as before. Therefore, from \eqref{omegahat} with \eqref{phat2l} and \eqref{what2l}, we have
\[
\begin{aligned}
\hat{\omega}_{2l}
&=
\frac{\tau^2\pi^2}{8l}\cdot\frac{3(b+cn)}{2F}\hat{\lambda}_{2l}.
\end{aligned}
\]
This shows that $\Omega_{2l}(\tau)=\frac{\tau^2\pi^2}{8l}\cdot\frac{3(b+cn)}{2F} > 0$ for all $\tau>0$.
\qed

\subsection{Proof for Theorem \ref{R5main}}\label{subsec:R5}

The distances between five evenly distributed regions are
\begin{equation}\label{disR5}
|x-y|=
\left\{
\begin{aligned}
&|0|=0,~\hspace{3mm}\mbox{if~}x= y,\\
&|1|=\frac{2\pi}{5},~\mbox{if~}x=y+1\pmod 5,\\
&|2|=\frac{4\pi}{5},~\mbox{if~}x=y+2\pmod 5,\\
&|3|=\frac{4\pi}{5},~\mbox{if~}x=y+3\pmod 5,\\
&|4|=\frac{2\pi}{5},~\mbox{if~}x=y+4\pmod 5,
\end{aligned}
\right.
\end{equation}
for $x, y\in\left\{0,1,2,3,4\right\}$. From \eqref{stpTheta} and \eqref{disR5} we see that
\begin{equation}\label{stp0p1R5}
\left\{\begin{aligned}
&\ol{p}_1 = \Theta + \frac{\tau}{2}|1| = \Theta + \frac{\tau\pi}{5}\\
&\ol{p}_2 = \Theta + \frac{\tau}{2}|2| = \Theta + \frac{2\tau\pi}{5}\\
&\ol{p}_3 = \Theta + \frac{\tau}{2}|3| = \Theta + \frac{2\tau\pi}{5} = \ol{p}_2\\
&\ol{p}_4 = \Theta + \frac{\tau}{2}|4| = \Theta + \frac{\tau\pi}{5} = \ol{p}_1.
\end{aligned}\right.
\end{equation}
We only need to discuss four effective frequencies $k=1,2,3$, and $4$. From \eqref{phat} and \eqref{stp0p1R5}, we have
\begin{equation}\label{phR5}
\left\{\begin{aligned}
&\hat{p}_1 = -\frac{c(3+\sqrt{5})}{10(2bF+c\Lambda)}\tau\pi\hat{\lambda}_1,\\
&\hat{p}_2 = -\frac{c(3-\sqrt{5})}{10(2bF+c\Lambda)}\tau\pi\hat{\lambda}_2,\\
&\hat{p}_3 = -\frac{c(3-\sqrt{5})}{10(2bF+c\Lambda)}\tau\pi\hat{\lambda}_3,\\
&\hat{p}_4 = -\frac{c(3+\sqrt{5})}{10(2bF+c\Lambda)}\tau\pi\hat{\lambda}_4.
\end{aligned}\right.
\end{equation}
Since
\[
\sum_{x=0}^4 |x-y| = 0+\frac{2\pi}{5}+\frac{4\pi}{5}+\frac{4\pi}{5}+\frac{2\pi}{5} = \frac{12\pi}{5},
\]
we see from \eqref{defTheta} that 
\begin{equation}\label{ThThbR5}
\Theta(\tau)  = \ol{\Theta}+\frac{6c\ol{\lambda}\tau\pi}{5(2bF+c\Lambda)}.
\end{equation}
Then, from \eqref{what}, \eqref{stp0p1R5}, \eqref{phR5}, \eqref{ThThbR5}, we have 
\begin{equation}\label{whR5}
\left\{
\begin{aligned}
\hat{w}_1 &= \frac{b+cn}{F}\left\{\frac{(3+\sqrt{5})\ol{\Theta}}{5}\tau\pi
\right.\\
&\hspace{22mm}\left.+\left[\frac{6(3+\sqrt{5})c\ol{\lambda}}{25(2bF+c\Lambda)} - \frac{5+3\sqrt{5}}{50}\right]\tau^2\pi^2\right\}\hat{\lambda}_1\\
&\hspace{10mm}+\frac{(\ol{\phi}+\ol{\lambda})(b+cn)(3+\sqrt{5})^2c}{50F(2bF+c\Lambda)}\tau^2\pi^2\hat{\lambda}_1,\\
\hat{w}_2 &= \frac{b+cn}{F} \left\{\frac{3-\sqrt{5}}{5}\ol{\Theta}\tau\pi \right.\\
&\hspace{22mm}\left.+ \left[\frac{6(3-\sqrt{5})c\ol{\lambda}}{25(2bF+c\Lambda)}-\frac{5-3\sqrt{5}}{50}\right]\tau^2\pi^2\right\}\hat{\lambda}_2\\
&\hspace{10mm}-\frac{c(\ol{\phi}+\ol{\lambda})(b+cn)(3-\sqrt{5})^2}{50F(2bF+c\Lambda)}\tau^2\pi^2\hat{\lambda}_2\\
\hat{w}_3 &= \frac{b+cn}{F} \left\{\frac{3-\sqrt{5}}{5}\ol{\Theta}\tau\pi \right.\\
&\hspace{22mm}\left.+ \left[\frac{6(3-\sqrt{5})c\ol{\lambda}}{25(2bF+c\Lambda)}-\frac{5-3\sqrt{5}}{50}\right]\tau^2\pi^2\right\}\hat{\lambda}_3\\
&\hspace{10mm}-\frac{c(\ol{\phi}+\ol{\lambda})(b+cn)(3-\sqrt{5})^2}{50F(2bF+c\Lambda)}\tau^2\pi^2\hat{\lambda}_3 \\
\hat{w}_4 &= \frac{b+cn}{F}\left\{\frac{(3+\sqrt{5})\ol{\Theta}}{5}\tau\pi
\right.\\
&\hspace{22mm}\left.+\left[\frac{6(3+\sqrt{5})c\ol{\lambda}}{25(2bF+c\Lambda)} - \frac{5+3\sqrt{5}}{50}\right]\tau^2\pi^2\right\}\hat{\lambda}_1\\
&\hspace{10mm}+\frac{(\ol{\phi}+\ol{\lambda})(b+cn)(3+\sqrt{5})^2c}{50F(2bF+c\Lambda)}\tau^2\pi^2\hat{\lambda}_4.
\end{aligned}\right.
\end{equation}
By the same manner, from \eqref{omegahat}, \eqref{stp0p1R5}, \eqref{phR5}, \eqref{ThThbR5}, we have 
\begin{equation}\label{omegahatR5}
\left\{
\begin{aligned}
\hat{\omega}_1 &= \hat{w}_1 
-\frac{a}{F}\left[\frac{-3-\sqrt{5}}{10}\tau\pi\right]\hat{\lambda}_1\\
&\hspace{10mm}+\frac{b+cn}{2F}\left[-\frac{3+\sqrt{5}}{5}\ol{\Theta}\tau\pi - \left(\frac{6(3+\sqrt{5})c\ol{\lambda}}{25(2bF+c\Lambda)}+\frac{5+3\sqrt{5}}{50}\right)\tau^2\pi^2\right]\hat{\lambda}_1\\
&\hspace{10mm}-\frac{c\ol{\lambda}}{F^2}
\left[5\ol{\Theta}+\left(\frac{6c\ol{\lambda}}{2bF+c\Lambda}+\frac{6}{5}\right)\tau\pi\right]\cdot\frac{-3-\sqrt{5}}{10}\tau\pi\hat{\lambda}_1 \\
&\hspace{10mm}+\left[-\frac{a\Lambda}{F}+\frac{b\Lambda}{5F}\left\{5\ol{\Theta}+\left(\frac{6c\ol{\lambda}}{2bF+c\Lambda}+\frac{6}{5}\right)\tau\pi\right\}\right]\cdot\frac{-c(3+\sqrt{5})}{10(2bF+c\Lambda)}\tau\pi \hat{\lambda}_1,\\
\hat{\omega}_2 &= \hat{w}_2
-\frac{a}{F}\left[\frac{-3+\sqrt{5}}{10}\tau\pi\right]\hat{\lambda}_2\\
&\hspace{10mm}+\frac{b+cn}{2F}\left[\frac{-3+\sqrt{5}}{5}\left(\ol{\Theta}+\frac{6c\ol{\lambda}\tau\pi}{5(2bF+c\Lambda)}\right)\tau\pi + \frac{-5+3\sqrt{5}}{50}\tau^2\pi^2\right]\hat{\lambda}_2\\
&\hspace{10mm}-\frac{c\ol{\lambda}}{F^2}\left[5\ol{\Theta}+\left(\frac{6c\ol{\lambda}}{2bF+c\Lambda}+\frac{6}{5}\right)\tau\pi\right]\left[\frac{-3+\sqrt{5}}{10}\tau\pi\right]\hat{\lambda}_2\\
&\hspace{10mm}+\left[-\frac{a\Lambda}{F}+\frac{b\Lambda}{5F}\left\{5\ol{\Theta}+\left(\frac{6c\ol{\lambda}}{2bF+c\Lambda}+\frac{6}{5}\right)\tau\pi\right\}\right]\cdot\frac{c(-3+\sqrt{5})}{10(2bF+c\Lambda)}\tau\pi\hat{\lambda}_2,\\
\hat{\omega}_3 &= \hat{w}_3
-\frac{a}{F}\left[\frac{-3+\sqrt{5}}{10}\tau\pi\right]\hat{\lambda}_3\\
&\hspace{10mm}+\frac{b+cn}{2F}\left[\frac{-3+\sqrt{5}}{5}\left(\ol{\Theta}+\frac{6c\ol{\lambda}\tau\pi}{5(2bF+c\Lambda)}\right)\tau\pi + \frac{-5+3\sqrt{5}}{50}\tau^2\pi^2\right]\hat{\lambda}_3\\
&\hspace{10mm}-\frac{c\ol{\lambda}}{F^2}\left[5\ol{\Theta}+\left(\frac{6c\ol{\lambda}}{2bF+c\Lambda}+\frac{6}{5}\right)\tau\pi\right]\left[\frac{-3+\sqrt{5}}{10}\tau\pi\right]\hat{\lambda}_3\\
&\hspace{10mm}+\left[-\frac{a\Lambda}{F}+\frac{b\Lambda}{5F}\left\{5\ol{\Theta}+\left(\frac{6c\ol{\lambda}}{2bF+c\Lambda}+\frac{6}{5}\right)\tau\pi\right\}\right]\cdot\frac{c(-3+\sqrt{5})}{10(2bF+c\Lambda)}\tau\pi\hat{\lambda}_3,\\
\hat{\omega}_4 &= \hat{w}_4 
-\frac{a}{F}\left[\frac{-3-\sqrt{5}}{10}\tau\pi\right]\hat{\lambda}_4\\
&\hspace{10mm}+\frac{b+cn}{2F}\left[-\frac{3+\sqrt{5}}{5}\ol{\Theta}\tau\pi - \left(\frac{6(3+\sqrt{5})c\ol{\lambda}}{25(2bF+c\Lambda)}+\frac{5+3\sqrt{5}}{50}\right)\tau^2\pi^2\right]\hat{\lambda}_4\\
&\hspace{10mm}-\frac{c\ol{\lambda}}{F^2}
\left[5\ol{\Theta}+\left(\frac{6c\ol{\lambda}}{2bF+c\Lambda}+\frac{6}{5}\right)\tau\pi\right]\cdot\frac{-3-\sqrt{5}}{10}\tau\pi\hat{\lambda}_4 \\
&\hspace{10mm}+\left[-\frac{a\Lambda}{F}+\frac{b\Lambda}{5F}\left\{5\ol{\Theta}+\left(\frac{6c\ol{\lambda}}{2bF+c\Lambda}+\frac{6}{5}\right)\tau\pi\right\}\right]\cdot \frac{-c(3+\sqrt{5})}{10(2bF+c\Lambda)}\tau\pi \hat{\lambda}_4.
\end{aligned}\right.
\end{equation}
Combining \eqref{whR5} and \eqref{omegahatR5}, we get
\[
\begin{aligned}
&\Omega_1(=\Omega_4) = A^{(5)}\tau^2+B^{(5)}\tau,\\
&\Omega_2(=\Omega_3) = C^{(5)}\tau^2+D^{(5)}\tau,
\end{aligned}
\]
where
\begin{equation}\label{AtoDR5}
\left\{
\begin{aligned}
A^{(5)} &= 
\frac{b+cn}{F}\left[\frac{6(3+\sqrt{5})c\ol{\lambda}}{25(2bF+c\Lambda)} - \frac{5+3\sqrt{5}}{50}\right]\pi^2\\
&\hspace{5mm}-\frac{c(\ol{\phi}+\ol{\lambda})(b+cn)(3+\sqrt{5})^2}{50F(2bF+c\Lambda)}\pi^2 \\
&\hspace{5mm}-\frac{b+cn}{2F}\left(\frac{6(3+\sqrt{5})c\ol{\lambda}}{25(2bF+c\Lambda)}+\frac{5+3\sqrt{5}}{50}\right)\pi^2 \\
&\hspace{5mm}+\frac{c\ol{\lambda}}{F^2}\cdot\frac{12(c\Lambda+bF)}{5(2bF+c\Lambda)}\cdot
\frac{3+\sqrt{5}}{10}\pi^2\\
&\hspace{5mm}-\frac{b\Lambda}{5F}\cdot\frac{12(c\Lambda+bF)}{5(2bF+c\Lambda)}\cdot\frac{c(3+\sqrt{5})}{10(2bF+c\Lambda)}\pi^2,\\
B^{(5)} &= 
\frac{(b+cn)(3+\sqrt{5})\ol{\Theta}}{5F}\pi
+\frac{a(3+\sqrt{5})}{10F}\pi\\
&\hspace{5mm}-\frac{(b+cn)(3+\sqrt{5})\ol{\Theta}}{10F}\pi
+\frac{c\ol{\lambda}5(3+\sqrt{5})\ol{\Theta}}{10F^2}\pi\\
&\hspace{5mm}+\frac{ac\Lambda (3+\sqrt{5})}{10F(2bF+c\Lambda)}\pi
-\frac{bc\Lambda (3+\sqrt{5})\ol{\Theta}}{10F(2bF+c\Lambda)}\pi,\\
C^{(5)} &= 
\frac{b+cn}{F}\left[\frac{6(3-\sqrt{5})c\ol{\lambda}}{25(2bF+c\Lambda)} - \frac{5-3\sqrt{5}}{50}\right]\pi^2\\
&\hspace{5mm}-\frac{c(\ol{\phi}+\ol{\lambda})(b+cn)(3-\sqrt{5})^2}{50F(2bF+c\Lambda)}\pi^2 \\
&\hspace{5mm}-\frac{b+cn}{2F}\left(\frac{6(3-\sqrt{5})c\ol{\lambda}}{25(2bF+c\Lambda)}-\frac{5-3\sqrt{5}}{50}\right)\pi^2 \\
&\hspace{5mm}+\frac{c\ol{\lambda}}{F^2}\cdot\frac{12(c\Lambda+bF)}{5(2bF+c\Lambda)}\cdot
\frac{3-\sqrt{5}}{10}\pi^2\\
&\hspace{5mm}-\frac{b\Lambda}{5F}\cdot\frac{12(c\Lambda+bF)}{5(2bF+c\Lambda)}\cdot\frac{c(3-\sqrt{5})}{10(2bF+c\Lambda)}\pi^2,\\
D^{(5)} &= 
\frac{(b+cn)(3-\sqrt{5})\ol{\Theta}}{5F}\pi
+\frac{a(3-\sqrt{5})}{10F}\pi\\
&\hspace{5mm}-\frac{(b+cn)(3-\sqrt{5})\ol{\Theta}}{10F}\pi
+\frac{c\Lambda(3+\sqrt{5})\ol{\Theta}}{10F^2}\pi\\
&\hspace{5mm}+\frac{ac\Lambda (3-\sqrt{5})}{10F(2bF+c\Lambda)}\pi
-\frac{bc\Lambda (3-\sqrt{5})\ol{\Theta}}{10F(2bF+c\Lambda)}\pi.\\
\end{aligned}\right.
\end{equation}
We can show careful calculation that $A^{(5)} < 0$ and $B^{(5)}>0$. These facts show that $\Omega_1(=\Omega_4)$ is an upwardly convex quadratic function passing through $\tau=0$ and $\tau^{(5)}>0$. 

We can show that $D^{(5)}>0$ as well. However, for $C^{(5)}$, we see that its sign depends critically on the parameter $\ol{\phi}$. In fact, when $\ol{\phi}=0$, we can show by careful calculation that $C^{(5)}>0$. Thus, in this case, we see that the function $\Omega_2(=\Omega_3)$ goes through the origin and $\Omega_2(=\Omega_3)>0$ for all $\tau>0$. On the other hand, if the value of $\ol{\phi}$ is sufficiently large, it is clear that $C^{(5)}<0$ by the the third term on the right-hand side of $C^{(5)}$ in \eqref{AtoDR5}.

\qed

\ifx\undefined\bysame
\newcommand{\bysame}{\leavevmode\hbox to\leftmargin{\hrulefill\,\,}}
\fi

\end{document}